\documentclass[12pt]{article}
\usepackage{spie,epsfig,latexsym}

\newcommand{\Z}{\mathbf{Z}}
\newcommand{\C}{\mathbf{C}}

\newcommand{\F}{\mathbf{F}}

\newcommand{\ket}[1]{|#1\rangle}
\newcommand{\mat}[4]{\left(\begin{array}{rrrr}#1&#2\\#3&#4\end{array}\right)}

\newcommand{\qed}{$\Box$}

\newcommand{\nix}[1]{}

\begin{document}
\author{Andreas Klappenecker\\
\small Department of Mathematics, Texas A\&M University\\[-1ex]
\small College Station, TX, 77843-3368, USA\\[-1ex]
\small \texttt{andreask@math.tamu.edu}
}
\title{Wavelets and Wavelet Packets\\ on Quantum Computers}
\maketitle
\begin{abstract} 
  We show how periodized wavelet packet transforms and periodized wavelet
  transforms can be implemented on a quantum computer. Surprisingly, we find
  that the implementation of wavelet packet transforms is less costly than
  the implementation of wavelet transforms on a quantum computer.
\end{abstract}

\section{Introduction}
Let $H$ be a Hilbert space with orthonormal basis $\{ e_k\,|\, k\in \Z\}$.
Recall that we can construct an new basis of $H$ by the following splitting
trick.\cite{meyer92,daubechies} Let $(\alpha,\beta)$ be a QMF system in $\ell^2(\Z)$,
that is, 
$$
\{ T_{2k}\, \alpha, T_{2k}\, \beta\,|\, k\in \Z\} \quad\mbox{is an
  orthonormal basis of}\quad \ell^2(\Z),
$$
where $T_k$ denotes the translation operator $T_k (s_n)_n =(s_{n-k})_n.$
For simplicity, we will also assume that $\alpha, \beta$ are finitely
supported. Then we obtain a new orthonormal basis $\{ f_k\,|\, k\in \Z\}$ of $H$ by
defining 
$$ f_{2k} = \sum_{l\in \Z} \alpha_{2k-l}\, e_l,\qquad
 f_{2k+1} = \sum_{l\in \Z} \beta_{2k-l}\, e_l,
$$ 
where $\alpha=(\alpha_k)_k$ and $\beta = (\beta_k)_k$. 

This splitting trick is used several times in wavelet and wavelet packet
algorithms. We may start with, say, the basis $e_k=\delta_k$, $k\in\Z$, where
$\delta_k$ is the sequence with value $1$ at $k$ and $0$ elsewhere.  Applying
the splitting trick, we obtain two closed subspaces $A$ and $D$ of $H$.
Namely, the closed subspace $A$ of $H$ generated by $\{f_{2k}\,|\,k\in \Z\}$, 
and the closed subspace $D$ of $H$ generated by $\{ f_{2k+1}\,|\,k\in
\Z\}$. We may split $A$ or $D$ as well. After a finite number of splitting
steps, we have a new basis of $H$.  The advantage is that the coordinate
change can be computed rather rapidly with quadrature mirror filter banks.

On a classical computer one tends to minimize the number of splitting steps,
mainly to reduce the cost of the computation. The wavelet algorithms split
only the spaces $A$ again. As a result, we obtain an algorithm with linear
complexity on a classical computer. The wavelet packet algorithms always split
both spaces $A$ and $D$, leading to $O(n\log n)$ operations on a classical
computer. Surprisingly, we will see that the implementation of wavelet packet
algorithms is less costly on a quantum computer than the implementation of
wavelet algorithms. 

In the next two sections we give a more or less self-contained introduction to
quantum circuits. Then we show how the Walsh-Hadamard transform can be
realized on a quantum computer. This is the simplest version of a wavelet
packet transform.  The rough architecture of wavelet packet and wavelet
transforms is described in section~5. We give a realization of the splitting
step in section~6.

\textit{Notation.} We denote by $\ell^2(\Z/N\Z)$ the complex vector space
$\C^N$ equipped with the usual inner product. This vector space should be
thought of as a periodized version of $\ell^2(\Z)$. Thus a vector in 
$\ell^2(\Z/N\Z)$ is given by a complex-valued sequence $(s_n)_{n\in \Z/N\Z}$. 

\section{Quantum Gates}
A quantum bit, or shortly qubit, takes a value in the complex two-dimensional
vector space $\C^2$. The standard basis of $\C^2$ is given by two orthogonal
vectors, denoted by $\ket{0}$ and $\ket{1}$. Dirac's ``ket'' notation is
traditionally used to describe the state of a quantum system. The labeling is
chosen to resemble the values $0$ and $1$ of a classical bit. However, the
state of a qubit can be described by a complex linear combination
$a\ket{0}+b\ket{1}$, unlike the classical case.

In classical computation, all operations on a single bit are given by the
identity mapping and the \textit{not} operation. In the quantum world, the
state of a qubit can be transformed by a unitary operation. This includes the
\textit{not} operation $U_{X} \ket{0} =\ket{1}$, $U_{X} \ket{1} =\ket{0}$, but
also many others. We write qubits as column vectors, and operate on these
vectors by left multiplication with unitary matrices. We express single bit
operations with respect to the basis $\{\ket{0},
\ket{1}\}$ unless otherwise specified.
Some of the operations
that will be used in the following can be described by the action of the
matrices:
$$
X =\mat{0}{1}{1}{0}, \qquad H=\frac{1}{\sqrt{2}}\mat{1}{1}{1}{-1}, \qquad
R(\theta) = \mat{\cos \theta}{\sin \theta}{-\sin \theta}{\cos \theta}.$$
The
matrix $X$ corresponds to the \textit{not} operation $U_{X}$. The operation
$U_H$, corresponding to the left multiplication by $H$, acts on the
classical states by
$$
U_H \ket{0} = \frac{1}{\sqrt{2}}\ket{0}+ \frac{1}{\sqrt{2}}\ket{1}, \qquad
U_H\ket{1} = \frac{1}{\sqrt{2}}\ket{0}- \frac{1}{\sqrt{2}}\ket{1}.
$$
The operator $U_H$ has no classical counterpart.

A finite collection of qubits is called a quantum register, or simply
register. The state of a register consisting of $n$ qubits can be described by
an element of the $(n-1)$-fold tensor product $V=\C^2\otimes \C^2 \otimes
\cdots \otimes \C^2$, a complex vector space of dimension $2^n$. This
remarkable property follows from the fundamental principle of quantum physics,
which asserts that the joint state of two quantum systems is the tensor product
of their individual quantum state spaces.

Consider a register with two quantum bits. A state of this register can be
expressed with respect to the basis $B=\left(\, \ket{0}\!\otimes\!\ket{0},
  \ket{0}\!\otimes\!\ket{1}, \ket{1}\!\otimes\!\ket{0},
  \ket{1}\!\otimes\!\ket{1} \,\right)$.  
Suppose we apply the single bit operation $U_H$ on the second (rightmost) qubit.
This operation acts on the state vector of the register 
by left multiplication with the matrix
$$I_2\otimes H = \frac{1}{\sqrt{2}}\left( 
\begin{array}{rrrr}
1 & 1 & 0 & 0 \\
1 & -1& 0 & 0 \\
0 & 0 & 1 & 1 \\
0 & 0 & 1 & -1
\end{array}
\right).$$
The single bit operations are considered as elementary operation on a
quantum computer. On a classical computer it can be a formidable task to
simulate these elementary operations, especially if the register consists
of a large number of qubits.

The notation of elements in the vector space $V$ is a little bit cumbersome.
We want to write for example $\ket{1} \otimes \ket{0} \otimes \ket{0}$ more
compactly as $\ket{100}$. For that reason, we fix an orthonormal basis of $\C^{2^n}$, and denote the basis elements by $\ket{x}$, where $x$ is a binary
vector in $\F_2^n$. We map the vector space
$V$ isomorphically onto $\C^{2^n}$ by
$$\ket{a_{n-1}}\otimes \cdots \otimes \ket{a_{1}}\otimes \ket{a_0} \longmapsto
\ket{a_{n-1}\dots a_1 a_0},$$
where the $a_i$ are elements of the finite field $\F_2$. We will refer to
$a_0$ as the lowest significant bit, and to $a_{n-1}$ as the highest
significant bit. We apply this isomorphism without further notice whenever it
is convenient. 

Apart from the single bit operations, we need operations manipulating several
qubits. The \textit{controlled not} gate manipulates two qubits, and is also
considered as an elementary operation in quantum computing. The controlled not
gate corresponds to a reversible version of the classical \textit{xor} gate.
It operates on the basis states by 
$$ 
\begin{array}{rrr}
\ket{0}\otimes\ket{0} & \mapsto & \ket{0}\otimes\ket{0}\\
\ket{0}\otimes\ket{1} & \mapsto & \ket{0}\otimes\ket{1}\\
\ket{1}\otimes\ket{0} & \mapsto & \ket{1}\otimes\ket{1}\\
\ket{1}\otimes\ket{1} & \mapsto & \ket{1}\otimes\ket{0}\\
\end{array}
\quad\mbox{or}\quad
\begin{array}{rrr}
\ket{00} & \mapsto & \ket{00}\\
\ket{01} & \mapsto & \ket{01}\\
\ket{10} & \mapsto & \ket{11}\\
\ket{11} & \mapsto & \ket{10}\\
\end{array}
$$
If the highest significant bit is $1$, then the state of the lowest
significant bit is flipped, that is, $\ket{a_1\, a_0} \mapsto \ket{a_1\,
  a_1\oplus a_0}$. We refer to the most significant bit as the \textit{control bit} and to the
least significant bit as the \textit{target bit} of the controlled not gate.
More generally, we can take two different qubits, take one as a control bit
and the other as a target bit. This way we obtain a controlled not gate on $n$ 
quantum bits. We will see some examples shortly.

The single bit gates and the controlled not bit gates are universal in the
sense that any unitary operator in $\C^{2^n}$ can be realized by a composition
of these gates\cite{BBC95}.

\section{Quantum Circuits}
Feynman introduced a graphical notation for quantum
gates\cite{feynman82,feynman96,BBC95,hoyer97}. It is convenient to specify
simple quantum circuits in this notation. The not gate $U_X$ with input $A$
and output $A'$ is denoted by 
\begin{center}
\epsfig{file=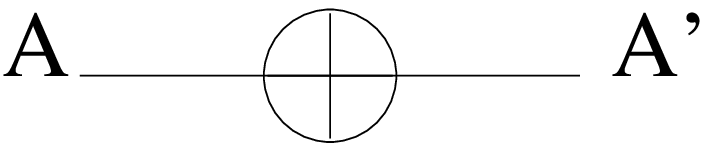,width=4cm}
\end{center}
A more general single bit gate $U_M$ is represented by
\begin{center}
\epsfig{file=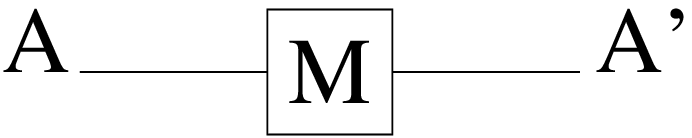,width=4cm}
\end{center}
A conditional not gate with two input $A$ and $B$ is denoted in the follwing way:
\begin{center}
\epsfig{file=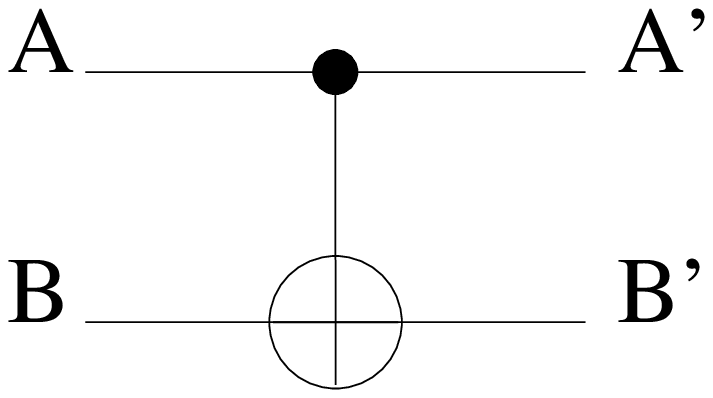,width=4cm}
\end{center}
The input $A$ controls the not operation on $B$. We adopt the convention that
the upper bit $A$ in the graphical notation corresponds to the most
significant (leftmost) bit, and that the lower bit $B$ corresponds to the
least significant bit. Thus, the conditional not gate shown above operates as
described in the previous section.

Similarly, we may apply a not operation on the lowest significant bit if
the highest significant bit is zero. The control bit is then given by a
non-filled circle in the graphical notation:
\begin{center}
\epsfig{file=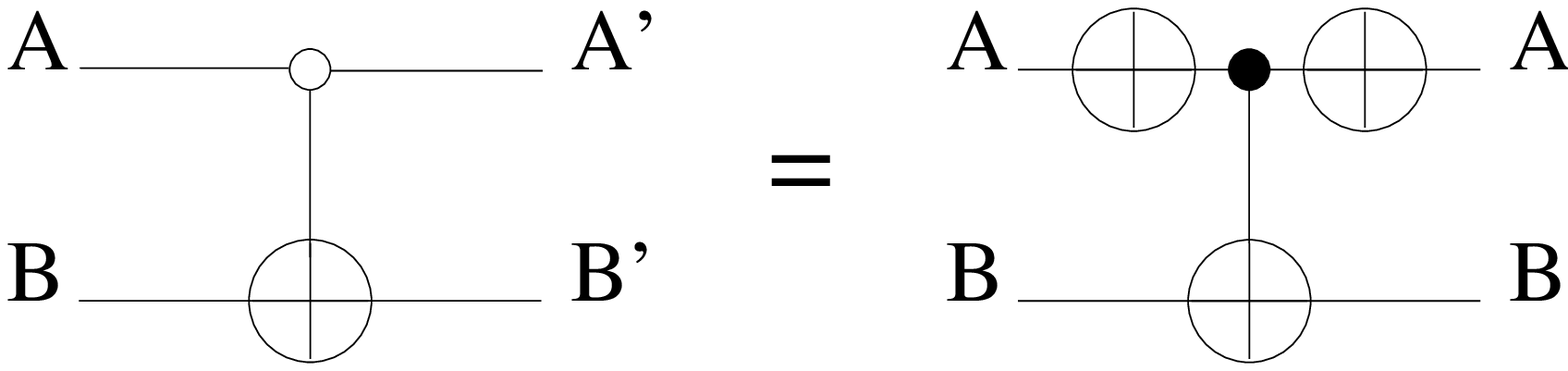,width=9cm}
\vspace*{-0.5cm}
\end{center}
In other words, the behaviour of this gate can be described by the map
$$
\begin{array}{rrr}
\ket{00}&\mapsto&\ket{01}\\
\ket{01}&\mapsto&\ket{00}\\
\ket{10}&\mapsto&\ket{10}\\ 
\ket{11}&\mapsto&\ket{11}
\end{array}\qquad\qquad
\overline{\mbox{C}}\mbox{NOT}:=\left(
\begin{array}{rrrr}
0&1&0&0\\
1&0&0&0\\
0&0&1&0\\
0&0&0&1
\end{array}
\right).
$$

As a convenient shorthand, we will use multiply conditioned single bit gates. 
For example, the following gate applies the single bit operation $U_M$ on the
least significant bit, if the higher significant bits $A_1,A_2,A_3$ are
$1,0,1$ respectively. 
\begin{center}
\epsfig{file=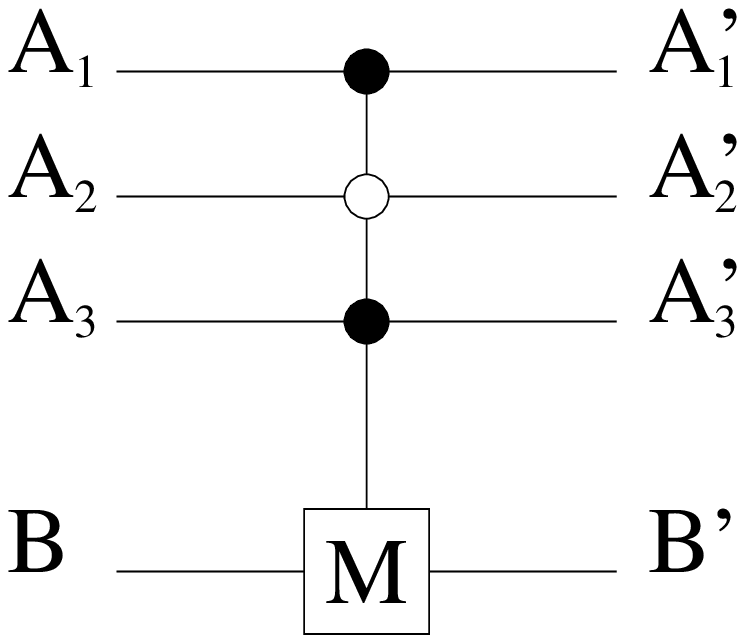,width=4cm}
\vspace*{-0.5cm}
\end{center}
It is well-known that such a multiply conditioned gate can be contructed with 
$\Theta(n)$ elementary gates, if we allow one additional qubit for
temporary calculations.\cite{BBC95} We will always assume that enough
additional (ancilla) qubits are available. 

\section{The Walsh-Hadamard Transform}
The Walsh-Hadamard transform is a simple example of a wavelet packet
transform. The quantum circuit is particularly simple in this case, yet it
shows some features of more complex wavelet packet transforms.

The Walsh-Hadamard transform $H_{2^n}$ for vectors of length $2^n$ can be
defined inductively by 
$$
H_2=H=\frac{1}{\sqrt{2}}\mat{1}{1}{1}{-1}, \qquad \mbox{and} \qquad H_{2^n} =
H_2\otimes H_{2^{n-1}}\quad \mbox{for}\quad n\ge 2.
$$
It is well-known \cite[p.~422]{macwilliams77} that $H_{2^n}$ can be factored as follows:
\begin{equation}\label{fht}
H_{2^n} = M_{2^n}^{(n)}M_{2^n}^{(n-1)}\cdots M_{2^n}^{(1)},
\qquad\mbox{where}\qquad M_{2^n}^{(k)} = I_{2^{n-k}} \otimes H \otimes
I_{2^{k-1}},
\end{equation}
and $I_n$ is an $n\times n$ unit matrix. We can directly translate this
factorization into a quantum circuit. Equation~(\ref{fht}) says that we have to
apply $U_H$ on each qubit, starting with the least significant bit. 
The following figure shows this circuit for signals of length 8: 
\begin{center}
\epsfig{file=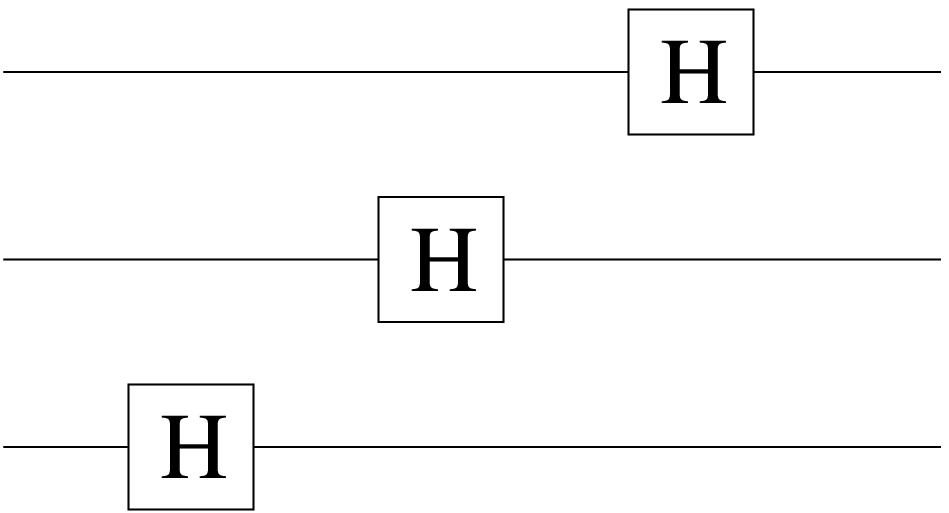,width=7cm}
\vspace*{-0.5cm}
\end{center}
The circuit is read like a musical score from left to right. We merely need
three elementary operations on a quantum computer. This circuit realizes the
matrix product below (which should be read from right to left, since we agreed
to act by left multiplication on the state vector of a quantum register):
\begin{small}
$$
\frac{1}{\sqrt{8}}
\left(
\begin{array}{rrrrrrrr}
1&.&.&.&1&.&.&.\\
.&1&.&.&.&1&.&.\\
.&.&1&.&.&.&1&.\\
.&.&.&1&.&.&.&1\\
1&.&.&.&-&.&.&.\\
.&1&.&.&.&-&.&.\\
.&.&1&.&.&.&-&.\\
.&.&.&1&.&.&.&-\\
\end{array}
\right)
\left(
\begin{array}{rrrrrrrr}
1&.&1&.&.&.&.&.\\
.&1&.&1&.&.&.&.\\
1&.&-&.&.&.&.&.\\
.&1&.&-&.&.&.&.\\
.&.&.&.&1&.&1&.\\
.&.&.&.&.&1&.&1\\
.&.&.&.&1&.&-&.\\
.&.&.&.&.&1&.&-\\
\end{array}
\right)
\left(
\begin{array}{rrrrrrrr}
1&1&.&.&.&.&.&.\\
1&-&.&.&.&.&.&.\\
.&.&1&1&.&.&.&.\\
.&.&1&-&.&.&.&.\\
.&.&.&.&1&1&.&.\\
.&.&.&.&1&-&.&.\\
.&.&.&.&.&.&1&1\\
.&.&.&.&.&.&1&-\\
\end{array}
\right)
$$
\end{small}
The dots denote $0$ and $-$ is short for $-1$.

We obtain a considerable speedup for larger values of $n$. While about $n2^n$
operations are needed on a classical computer, only $n$ elementary operations
are needed on a quantum computer. The fast Walsh-Hadamard transform is an
essential ingredient of Simon's algorithm\cite{Sim94}.

\section{Wavelet Packet Transforms}
The state space of a quantum computer is large, but finite dimensional. For
simplicity, I decided to discuss only periodic (also known as cyclic) wavelet
packet transforms. Thus, signals and filters are regarded as periodic
sequences. The benefit is that we obtain rather simple quantum circuits. Other
methods of border treatment will in general increase the complexity of the
implementation. Since adding a single quantum bit allows us to double the
signal length, it is easy to emulate non-periodic versions by choosing a large
signal period.

Let us recast the splitting trick for finite dimensions.
Let $H$ a finite dimensional Hilbert space with orthonormal basis 
$\{ e_k\,|\, k\in \Z/2N\Z\}$. Let $(\alpha,\beta)$ a QMF system of
$\ell^2(\Z/2N\Z)$, that is,
$$
\left\{ T_{2k} \alpha, T_{2k}\beta\,|\, k\in [0\!:\!N-1]\, \right\} \quad
\mbox{is an orthonormal basis of}\quad \ell^2(\Z/2N\Z),$$
where $T_k\,
e_n=e_{k+n}$ for all $n\in \Z/2N\Z$. 

We obtain a new orthonormal basis $\{
f_k\,|\, k\in \Z/2N\Z\}$ of $H$ by defining
$$ f_{2k} = \sum_n \alpha_{n-2k}\, e_n,\qquad 
f_{2k+1} = \sum_n \beta_{n-2k}\,e_n,$$
where $\alpha=(\alpha_k)$ and $\beta=(\beta_k)$. The proof is an immediate
consequence of the definitions. 

Let me illustrate this for the case $N=4$, that is, the dimension of $H$ is
$8$. According to the splitting trick, we obtain the following base change
matrix from $(e_n)$ to $(f_n)$: 
$$
\left(
\begin{array}{rrrrrrrr}
\alpha_0&\beta_0&.&.&.&.&\alpha_2&\beta_2\\
\alpha_1&\beta_1&.&.&.&.&\alpha_3&\beta_3\\
\alpha_2&\beta_2&\alpha_0&\beta_0&.&.&.&.\\
\alpha_3&\beta_3&\alpha_1&\beta_1&.&.&.&.\\
    .&.&\alpha_2&\beta_2&\alpha_0&\beta_0&.&.\\
    .&.&\alpha_3&\beta_3&\alpha_1&\beta_1&.&.\\
        .&.&.&.&\alpha_2&\beta_2&\alpha_0&\beta_0\\
        .&.&.&.&\alpha_3&\beta_3&\alpha_1&\beta_1\\
\end{array}
\right)
$$

Assume that we are given the signal as a component vector with respect to the
base $(e_n)$. Then the splitting step corresponds to left multiplication with
the matrix 
$$
\left(
\begin{array}{rrrrrrrr}
\overline{\alpha}_0&\overline{\alpha}_1&\overline{\alpha}_2&\overline{\alpha}_3&.&.&.&.\\
\overline{\beta}_0&\overline{\beta}_1&\overline{\beta}_2&\overline{\beta}_3&.&.&.&.\\
.&.&\overline{\alpha}_0&\overline{\alpha}_1&\overline{\alpha}_2&\overline{\alpha}_3&.&.\\
.&.&\overline{\beta}_0&\overline{\beta}_1&\overline{\beta}_2&\overline{\beta}_3&.&.\\
.&.&.&.&\overline{\alpha_0}&\overline{\alpha}_1&\overline{\alpha}_2&\overline{\alpha}_3\\
.&.&.&.&\overline{\beta_0}&\overline{\beta}_1&\overline{\beta}_2&\overline{\beta}_3\\
\overline{\alpha}_2&\overline{\alpha}_3&.&.&.&.&\overline{\alpha}_0&\overline{\alpha}_1\\
\overline{\beta}_2&\overline{\beta}_3&.&.&.&.&\overline{\beta}_0&\overline{\beta}_1\\
\end{array}
\right)
$$
Setting $\alpha_0=\alpha_1=\beta_0=1/\sqrt{2}$, $\beta_1=-1/\sqrt{2}$, and
all other coefficients zero, we recognize the first splitting step of the
Walsh-Hadamard transform. The coefficients with even index contain the sum
(approximation) and the coefficients with odd index contain the difference
(detail).

We will see in the next section how such splitting matrices can be
implemented.  The next observation is essential for wavelet packet algorithms.
Suppose we want to split both spaces $A=\mbox{span} \{\,T_k\, \alpha \,|\,
k=0,2,4,6\,\}$ and $D=\mbox{span} \{\,T_k\, \beta \,|\, k=0,2,4,6\,\}$. Then
we observe that \emph{both} splitting steps can be realized by the following
tensor product of matrices:
$$
\left(
\begin{array}{rrrr}
\overline{\alpha}_0&\overline{\alpha}_1&\overline{\alpha}_2&\overline{\alpha}_3\\
\overline{\beta}_0&\overline{\beta}_1&\overline{\beta}_2&\overline{\beta}_3\\
\overline{\alpha}_2&\overline{\alpha}_3&\overline{\alpha}_0&\overline{\alpha}_1\\
\overline{\beta}_2&\overline{\beta}_3&\overline{\beta}_0&\overline{\beta}_1\\
\end{array}
\right)
\otimes I_2 = 
\left(
\begin{array}{rrrrrrrr}
\overline{\alpha}_0&.&\overline{\alpha}_1&.&\overline{\alpha}_2&.&\overline{\alpha}_3&.\\
.&\overline{\alpha}_0&.&\overline{\alpha}_1&.&\overline{\alpha}_2&.&\overline{\alpha}_3\\
\overline{\beta}_0&.&\overline{\beta}_1&.&\overline{\beta}_2&.&\overline{\beta}_3&.\\
.&\overline{\beta}_0&.&\overline{\beta}_1&.&\overline{\beta}_2&.&\overline{\beta}_3\\
\overline{\alpha}_2&.&\overline{\alpha}_3&.&\overline{\alpha}_0&.&\overline{\alpha}_1&.\\
.&\overline{\alpha}_2&.&\overline{\alpha}_3&.&\overline{\alpha}_0&.&\overline{\alpha}_1\\
\overline{\beta}_2&.&\overline{\beta}_3&.&\overline{\beta}_0&.&\overline{\beta}_1&.\\
.&\overline{\beta}_2&.&\overline{\beta}_3&.&\overline{\beta}_0&.&\overline{\beta}_1\\
\end{array}
\right).
$$

Assume that \texttt{SPLIT}n is a quantum circuit realizing a splitting step for
$n$-dimensional signals. Then the quantum circuit for a wavelet packet tree of
depth two is shown below: 
\begin{center}
\epsfig{file=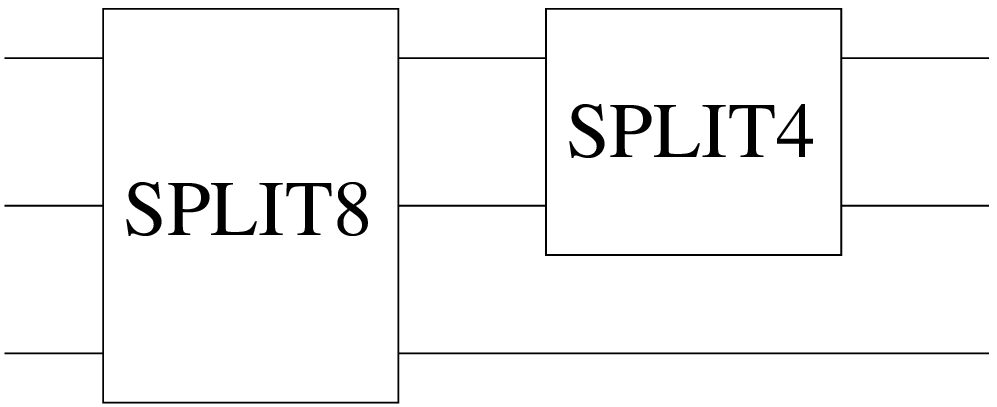, width=8cm}
\end{center}

If we decide to realize only a splitting of the space $A$, then we have to
condition \texttt{SPLIT4} in the following way:
\begin{center}
\epsfig{file=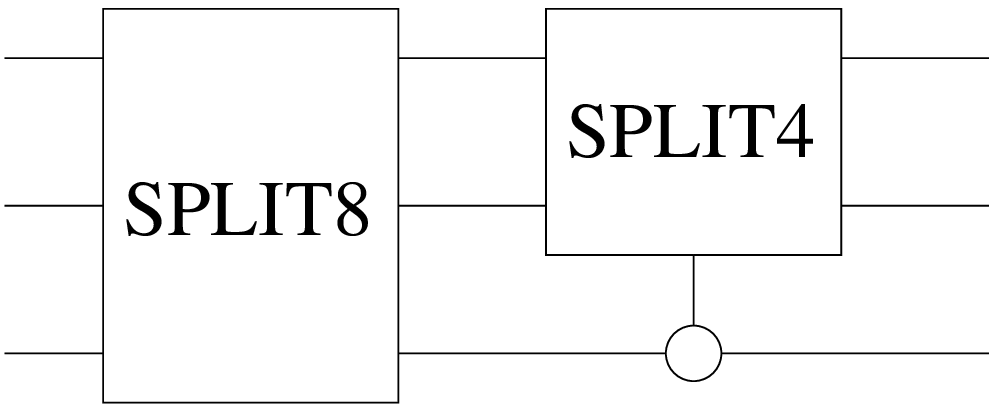, width=8cm}
\end{center}
The conditioning means that we have to add this further condition to all gates
(which is rather costly), or we have to prevent the circuit from changing the
input in case the condition is not satisfied. There are several different ways
to do that. I will indicate in the next section how the splitting steps can be
conditioned.

\section{The Splitting Step}
The base change operator $O$ from $(e_n)$ to $(f_n)$ satisfies 
a remarkable property: 
$$
f_0 = O e_0, \qquad f_1 = O e_1,$$
and $O$ commutes with the even
translations $OT_2=T_2O$. This follows from the property $T_{2k} f_0 = f_{2k}$
and $T_{2k} f_1 = f_{2k+1}$ of the basis $(f_n)$. On the other hand, it is
clear that any unitary operator on $H$ commuting with $T_2$ corresponds to a
spitting trick construction. 

Let us define a few simple operators commuting with $T_2$.  Denote by $M$ a
unitary $2\times 2$-matrix. Then the operator $O_M$, given by the matrix
$I_N\otimes M$ with respect to the basis $(e_n)$, is easily seen to commute
with $T_2$. We refer to $O_M$ as a local rotation operator.  Note that the
translation operator $T_1$ commutes with $T_2$ as well. We will construct our
splitting steps as products of translation operators and local rotation
operators.

\paragraph{Case $\mathbf{N=2^n}$.}

A circuit for the local rotation operator $O_M$ is simply given by a single
bit operation on the lowest significant input bit. We merely need to show how
a circuit for $\ket{m} \mapsto \ket{m+1 \bmod 2^n}$ can be constructed.  In
binary representation, this mapping can be specified in terms of the following
operations in $\F_2$:

$$\ket{a_{n-1}\dots a_1a_0} \longmapsto\ket{b_{n-1}\dots b_1b_0},
\qquad a_i, b_i \in \F_2,$$ 
with 
$$ 
\begin{array}{rcll}
b_0 &=& a_0 + 1\\
b_1 &=& a_1 + a_0\\
b_2 &=& a_2 + c_1, &\quad\mbox{where}\quad c_1 = a_1a_0,\\
b_{i} &=& a_{i}+ c_{i-1}, &\quad\mbox{where}\quad c_{i-1} = a_{i-1}c_{i-2},
\end{array}
$$
for $3\le i\le n-1$. Allowing additional qubits for the calculation of the
carries $c_i$, we obtain a particular simple implementation. 
Calculating the $c_i$'s and then the $b_i$'s, we obtain the following circuit
for the case $n=4$:
\begin{center}
\epsfig{file=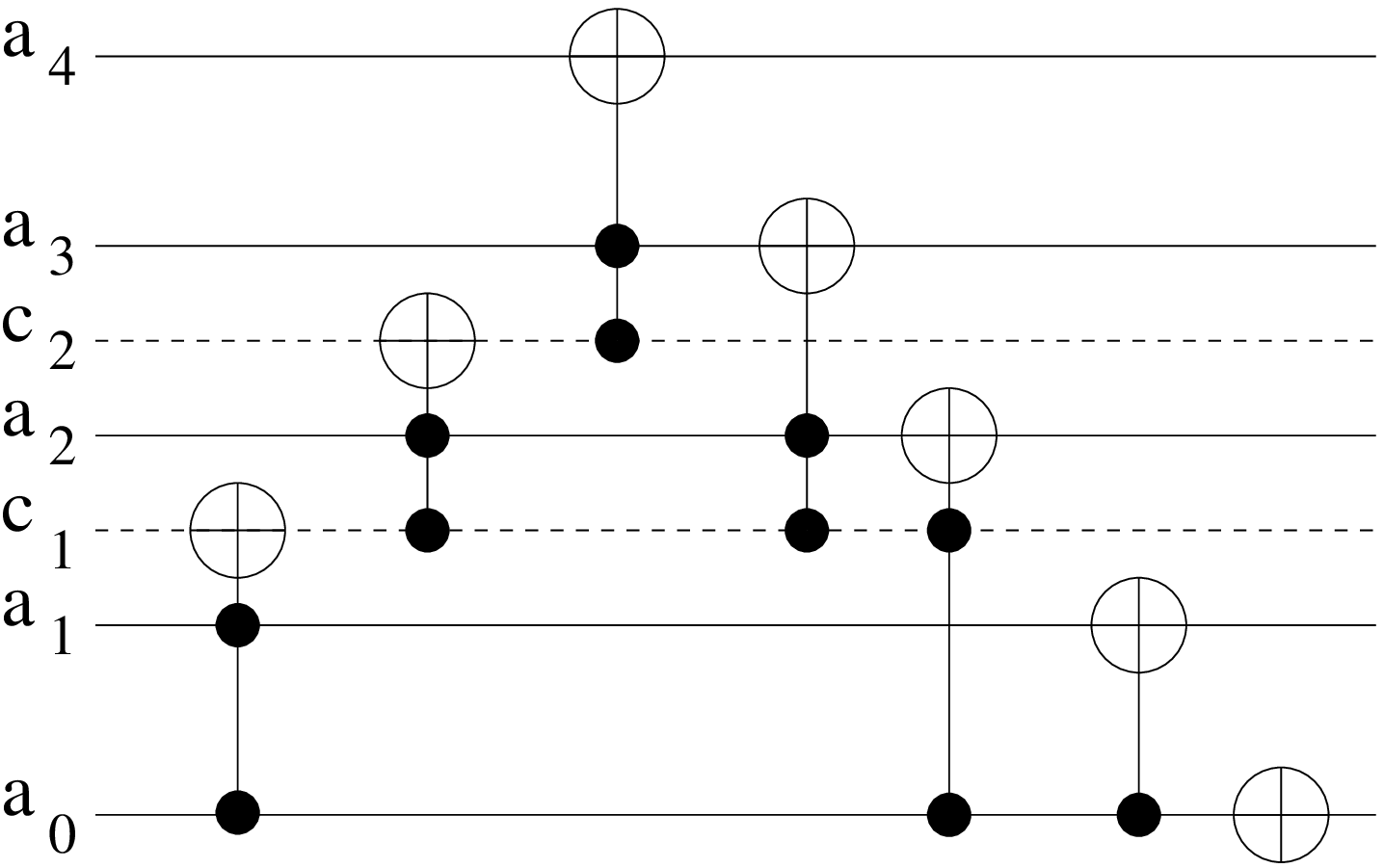, width=8cm}
\end{center}
The qubits for the carries $c_0$ and $c_1$ are initally prepared in the state
$\ket{0}$. Cleaning up operations of these ancilla bits are not shown.

\begin{lemma}
  The translation operations $\ket{m} \mapsto \ket{m+1 \bmod 2^n}$ and
  $\ket{m} \mapsto \ket{m-1 \bmod 2^n}$ can be implemented with $O(n)$
  elementary quantum gates, if additional qubits are allowed for temporary
  calculations.
\end{lemma}
\proof For $n\ge 3$, we need $n-2$ Toffoli gates (conditional not gates with
two conditions) for the calculation of the carries, the same number is needed
to clean up the ancilla bits.  One conditional not gate, one not gate, and
$n-1$ Toffoli gates are needed for the calculations of the $b_i$'s.  How
Toffoli gates can be expressed in terms of elementary gates is described in
Barenco et.~al.\cite{BBC95} Running the circuit backwards yields a circuit
for $\ket{m} \mapsto \ket{m-1 \bmod 2^n}$.~\qed

\textbf{Remark.} For wavelet algorithms we need a conditioned form of
circuits. The local rotation operation is simply a conditioned single bit
gate. It is well-known how to express such a gate in terms of elementary
gates\cite{BBC95}. Conditioning of the translation operations is also easy.
We just condition the gates with target $c_1$, $a_1$, and $a_0$. Conditioning
$c_1$ means that the carry does not ripple through, hence the higher
significant bits are not changed. The gates with target bits $a_1$ and $a_0$
need to be conditioned to prevent a distortion of the two least significant
bits.

\nix{
\begin{lemma}
Suppose we have a splitting step operating on $n_1$ qubits which is subject to
$n_2$-conditions. This circuit can be
implemented with $O(n_1+n_2)$ elementary gates, provided we allow ancilla
qubits for temporary calculations.
\end{lemma}
\proof Note that a conditional not gate with $n_2$-conditions can be
implemented with $O(n_2)$ operations. Let the target of this gate be an
ancilla qubit prepared in the state $\ket{0}$. Thus if the condition is
satisfied this ancilla bit will be flipped. According to the previous remark 
}

\paragraph{General Case.}
Some modifications are needed if we want to implement a splitting step for
signals of even length $2N$, $N$ not a power of two. The basic idea is to
embed $H$ into a complex vector space of dimension $2^n$, where $n :=
\lceil \log_2(2N)\rceil$. We assume that the input is given as a linear
combination of $\ket{0}, \dots, \ket{2N-1}$. We may still realize the local
rotation operator by applying a single bit operation on the least significant
qubit.

We realize the operation $\ket{m}\mapsto \ket{m+1 \bmod 2N}$ as follows.  Note
that $\ket{m}\mapsto \ket{m+1 \bmod 2^n}$ maps $\ket{m}$ to $\ket{m+1}$, for
$0\le m< 2N-1$, just in the way we want it. Only $\ket{2N-1}$ is mapped to
$\ket{2N}$ instead of $\ket{0}$. We can take care of this exception by setting
an additional qubit. Then we merely need to set all $1$'s in the binary
representation of $2N$ to 0. For example, if we want to realize the circuit
$\ket{m}\mapsto{m+1 \bmod 6}$, then we may take the circuit for $x\mapsto x+1
\bmod 8$ and modify it in the following way:
\begin{center}
\epsfig{file=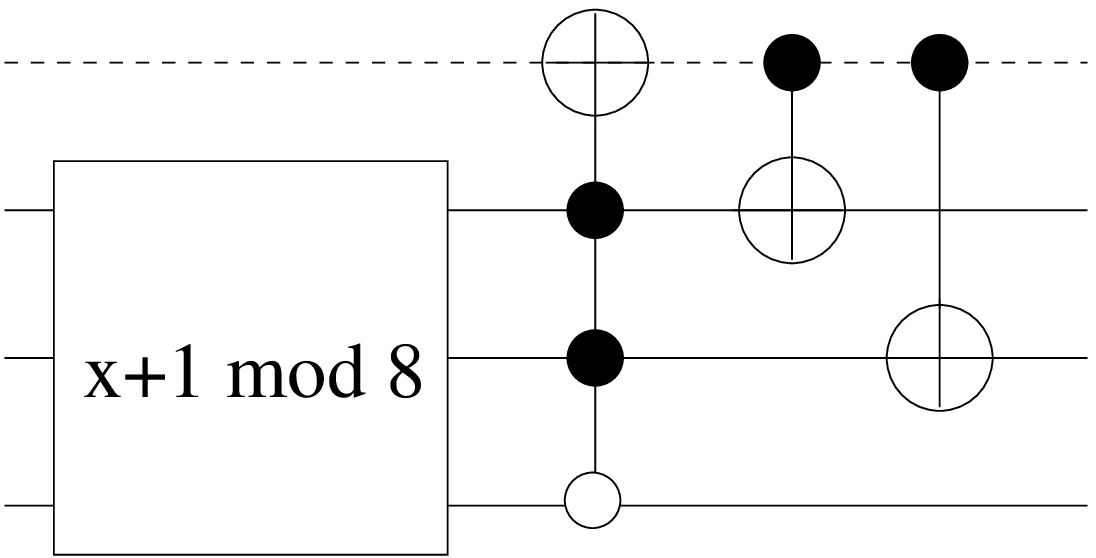, width=8cm}
\end{center}
Here it is again assumed that the additional qubit is initally prepared in the
state $\ket{0}$. Note that a $\ket{m}\mapsto\ket{m+1 \bmod 2N}$ gate can still
be implemented with $O(n)$ elementary gates.

\paragraph{Remarks:} 
\vspace{-2ex}
\begin{enumerate}
\item Our construction of splitting steps is motivated by the parametrization
  of Holschneider and Pinkall\cite{holschneider93,holschneiderbook} of QMF
  systems of $\ell^2(\Z)$.  They essentially showed that any finitely
  supported QMF system can be obtained by applying local rotation and
  translation operators on $(\delta_0,\delta_1)$.  Consequently, we can factor
  all QMF systems of $\ell^2(\Z/2N\Z)$ that are periodized versions of QMF
  systems of $\ell^2(\Z)$. See [\citenum{klappeneckerdiss,beth98}] for
  more details on this.
\item Note, however, that the parametrization is in general incomplete in the
  finite dimensional case.  This is easy to see by looking at the polyphase
  matrices of our QMF filters.  The determinant of a polyphase matrix obtained
  by our construction has trivial units in the group ring $\C[\Z/N\Z]$. 
\end{enumerate}

\section{Conclusion}
Suppose we want to compute either a wavelet or a wavelet packet transform on a
quantum computer.  Let us assume that each splitting step has the same number
$L$ of local rotation and translation operators. For example, this is the case
if we take a periodized version of a fixed QMF system of $\ell^2(\Z)$ at each
step.  Assuming that the signal can be represented by $n$ qubits, then we have
at most $n$ (conditioned) splitting circuits to realize each transform.  The
complexity of each conditioned splitting step is at most $O(Ln)$. Hence at
most $O(Ln^2)$ elementary gates are needed to realize the transform. Note that
wavelet packet algorithms need fewer operations than wavelet algorithms on a
quantum computer; this fact is hidden in the constant of the $O$-notation.

Let $N=2^n$ be the length of the input signal. On a classical computer we need
$O(N)$ operations for a wavelet transform and $O(N\log N)$ operations for
a wavelet packet transform. On a quantum computer we merely need $O(\log^2
N)$ elementary quantum gate operations for a wavelet or a wavelet packet
transform.

\paragraph{Acknowledgements.} I wish to thank Professor Thomas Beth and Markus
Grassl for introducing me to quantum computing.

\appendix
\section{Program}
I have written a small perl program to simulate simple quantum gates.
This toy is freely available from the following web site: 
\par 
\texttt{http://www.math.tamu.edu/\~{}Andreas.Klappenecker/}
\par
\noindent 
The purpose of this program is only educational. It is my belief that the
basic ideas of quantum circuits can be learned rather quickly. Playing around
with this toy program may help in this process.

\bibliography{wavelet,mathe,quantum}
\bibliographystyle{./spiebib}
\end{document}